# Multiple timescales in collective motion: daily and intraday upstream fish migration focusing on Feller condition

Hidekazu Yoshioka[1, *]

[1] Graduate School of Advanced Science and Technology, Japan Advanced Institute of Science and Technology, 1-1 Asahidai, Nomi, Ishikawa, Japan

* Corresponding author: yoshih@jaist.ac.jp, ORCID: 0000-0002-5293-3246

***Abstract***

Fish migration is a collective phenomenon that has multiple timescales, ranging from daily to intraday (hourly or even finer). We propose a unified mathematical approach using diffusion bridges, nonlinear stochastic differential equations with pinned initial and terminal conditions, to model both daily and intraday fish migration phenomena. Drift and diffusion coefficients of these bridges are determined based on time-dependent parameterized average and variance curves fitted against fish count data, with which the unique existence of their solutions is rigorously guaranteed. We show that sample paths of the diffusion bridges have qualitatively distinctive properties depending on the Feller condition, namely, the ratio between the sizes of diffusion and drift. Our application study about the juvenile upstream migration of *Plecoglossus altivelis altivelis* (Ayu) in Japan clarifies similarities and differences between daily and intraday migration phenomena. Particularly, we discuss that the daily and intraday fish count data correspond to distinctive Feller indices, showing that the former is qualitatively less randomized and intermittent. The results obtained in this study suggest that the Feller condition potentially serves as an effective tool for evaluating fish migration phenomena of Ayu across different timescales.

***Statements & Declarations***

**Acknowledgments:** The author would like to express his gratitude to the Japan Water Agency, Ibi River, and Nagara River General Management Office for providing valuable fish migration data of *Plecoglossus altivelis altivelis* at the Nagara River Estuary Barrage. The author is very grateful for the comments from the anonymous reviewers.

**Funding:** This study was supported by the Japan Science and Technology Agency (PRESTO No. JPMJPR24KE).

**Conflict of Interests:** The authors have no relevant financial or non-financial interests to disclose.

**Data Availability:** The data will be made available at reasonable request to the corresponding author.

**Declaration of Generative AI in Scientific Writing:** The authors did not use generative AI in the scientific writing of this manuscript.

**Contribution:** The corresponding author prepared all parts of this manuscript.

## 1. Introduction

### 1.1 Study background

Collective motion of biological organisms occurs in various spatial and temporal scales [1-3], and those found in earth environments, such as migration of fish, birds, and insects essentially drive ecosystems [4-6]. Fish migration is important for human societies because many migratory fish species are fishery resources for human lives [7-9]. In this study, we focus on fish migration phenomena because they are socially important and have many unresolved issues. Particularly, studies on fish migration with hourly or finer temporal resolution are scarce, possibly owing to technical and labor reasons.

A lot of studies focused on daily fish migration phenomena. Migration phenomena in daily timescales have been found to reflect water environmental changes due to anthropogenic [10,11] and climatic factors [12,13]. Daily fish location data contain effective information about long-distance migration along river reaches [14,15]. Mathematical modeling of daily fish migration has been conducted in several studies. Paíz et al. [16] developed a probabilistic model for predicting the timing of mass migration events. Wang and Huang [17] investigated the daily mortality and migration dynamics of sturgeon based on a convection-diffusion equation model. Lobyrev et al. [18] proposed a daily spatio-temporal stochastic model that accounts for age structure and growth dynamics. Migration dynamics of salmon in a stream network on daily or longer timescales have been analyzed using a habitat suitability model [19]. A machine-learning approach has been applied to simulate daily trajectories of sea capelin in three-dimensional oceanic fields [20]. Yoshioka [21] applied a memory-based stochastic process to daily fish migration data and evaluated the fluctuation and autocorrelation in the upstream migration of diadromous fish.

By contrast, few studies focused on intraday fish migration phenomena. Existing studies discussed microscopic migration behavior of fish, such as the nighttime spawning migration of shad [22], unsteady vertical swimming of sea trout in daytime that depends on fish size and water depth [23], upstream migration of trout influenced by a submerged weir whose water depth is controlled by hydropower generation [24], and intraday passage success evaluation of smolt salmon across a bypass channel installed in a hydropower plant [25]. Agent-based models that consider fish as active particles have also been applied to intraday and finer fish migration in water bodies and hydraulic structures [26-28]. Yoshioka [29] proposed a continuous-time stochastic model with pinned initial and terminal conditions for describing the juvenile upstream migration of the common diadromous fish *Plecoglossus altivelis altivelis* (Ayu) in Japan.

Thus, fish migration has been investigated by focusing on distinctive timescales; however, few studies concurrently investigated migrations under multiple timescales, hindering deeper understanding of this complex biological phenomenon. One unresolved issue is whether one mathematical model can be applied to both intraday and daily fish migration phenomena, where the former occurs within several hours, and the latter continues for more than a few months. In physics, problems with different timescales are often described by the same equation, such as the Navier–Stokes equations for incompressible fluid flows [30], although it is not for modeling fish migration. For collective motion of animals, social interactions among individuals occur in various timescales [31-33], but their unified descriptions have not been established.

Addressing the aforementioned multiscale modeling issue will deepen our understanding of fish migration as a complex and dynamic biological phenomenon. Applying a common mathematical model with properly scaled parameters and coefficients simultaneously to different migration regimes with distinctive timescales can effectively characterize theoretical similarities and differences in fish migration across different scales. This is the main motivation for this study.

### 1.2 Aim and contribution of this study

The aims of this study are twofold: to propose a unified mathematical description of seasonal fish migration for both daily and intraday timescales and to apply the model to real data. The core hypothesis behind the proposed approach is that daily and intraday fish count data can be described by a class of stochastic differential equations (SDEs) driven by Brownian motions [34] with different parameter values depending on the target timescale. The advantage of SDE is its ability to efficiently (both theoretically and computationally) model complex and nonlinear phenomena, while its limitation is the phenomenological nature, as unmodeled complexities are aggregated into the stochastic noise term. Applying the same SDE to migration phenomena would enable qualitative and quantitative comparison of their randomness by a common (normalized) standard, with which their similarities and differences can be understood clearly.

The proposed mathematical framework is a generalization of the Cox–Ingersoll–Ross (CIR) bridge proposed in Yoshioka [29], which is a diffusive SDE with a nonlinear and non-Lipschitz diffusion term describing unit-time fish count at a fixed observation point between each sunrise and sunset. This model is based on the classical CIR model serving as a minimal nonlinear SDE for nonnegative and continuous-time stochastic processes originally proposed for describing interest rate dynamics [35]. Advantages of the CIR model are its well-posedness (e.g., admits unique nonnegative solutions in a pathwise sense) and tractability owing to the explicit availability of major statistics such as cumulants and autocorrelation [e.g., 36,37]. CIR models have been widely used in applied study areas, such as option pricing [38], key mortality rates [39], high-resolution imaging [40], wind speed dynamics [41], and rainfall dynamics [42].

The CIR bridge extends the classical model with singular (i.e., blowing-up) drift and diffusion coefficients so that the model is defined in each daytime during which the fish migration is assumed to occur. The CIR bridge can be a minimal stochastic model that accounts for the nonnegativity, intermittency, and time constraints on when upstream migration occurs. Mathematically, the nonlinear diffusion term realizes these properties at once. Particularly, the singularity of coefficients enables us to model the intraday fish migration to occur only during daytime. Despite that it is a nonlinear model having singular coefficients, the CIR bridge inherits the well-posedness (under certain conditions; see **Section 2**) and analytical tractability with which model parameters can be efficiently identified from data. Another common advantage shared by the classical CIR model and CIR bridge is that they can describe both intermittent and non-intermittent sample paths by properly scaling drift and/or diffusion coefficients. Here, intermittency means that a nonnegative sample path hits the minimum value 0 in a finite time with probability 1 at each time. In the classical CIR model, the necessary and sufficient condition to judge intermittency is the Feller condition, which is qualitatively the ratio between variance and the square of average (see **Section 2.3** in this paper), and its sample path becomes intermittent if and only if the diffusion (or equivalently nonlinearity) is sufficiently large (Proposition 1.2.15 in Alfonsi [36]). The Feller condition is said to be satisfied (resp., violated) if the model gives non-intermittent (intermittent) sample paths. The CIR bridge was found to inherit the intermittency as well as Feller condition with slight modifications, where the intraday fish count data have been judged to be intermittent [29].

Interestingly, financial and economic application studies based on the CIR and related models often assume that they are not intermittent, although some real data deny this assumption [43,44], and there are analytical results without the Feller condition [45-47]. Indeed, non-intermittent cases are easier to address from both theoretical and computational standpoints [48]. Because intermittency means difficulties in prediction with coarse but practical observation schemes according to the statistical analysis [49], a theoretical interest is whether daily fish migration phenomena are intermittent or not, and how the intermittency is different between distinctive timescales.

A concept behind applying CIR bridges to daily data is that seasonal fish migration starts and ends in some time interval, ranging from several weeks to months each year [21,50-54]. This finding motivates us to apply CIR bridges to daily fish migration phenomena by properly scaling that for intraday. In our context, the Feller condition quantifies the predictability of migration phenomena depending on timescales. To the best of the author's knowledge, such a mathematical approach has not been widely explored.

A critical difference between the present and previous studies [21,29,49,55] is that the former considers both daily and intraday migration phenomena while the latter with previous approaches. Another difference between them is the modeling strategies, where the present study models a CIR bridge starting from the average and variance curves, while the latter starts from drift and diffusion coefficients. The present approach can directly connect the statistics and model irrespective of timescales.

We apply CIR bridges along with their Feller condition to the intraday (10-min) and daily fish count data of juvenile upstream migration of Ayu near the mouth of the Nagara River in Japan, which is a valuable site where both daily and intraday fish count data are available. Ayu is a model fish species without age structures, and its life cycle is typically closed in one year, which is the main reason for focusing on this fish species in recent studies [e.g., 21,29,49]. Coefficients of the CIR bridge are designed based on time-dependent parameterized average and variance curves inspired from real data. This kind of design method of SDEs is similar to diffusion models [56], but ours seems to be theoretically simpler because it does not aim at iteratively learning data. We show that the intraday and daily fish migration have significantly different randomness in view of the Feller condition. We show that both cases violate the Feller condition and the corresponding sample paths are truly intermittent, and the intermittency and randomness are weaker in the daily case than in the intraday one. Consequently, this study contributes to the development and application of a unified mathematical method for collective motion of biological organisms, focusing on Ayu as a case study.

### 1.3 Structure of the remainder of this paper

**Section 2** presents our CIR bridge and explains the proposed approach for dealing with daily and intraday migration phenomena. **Section 3** applies the proposed model to real data. **Section 4** concludes the study and presents future perspectives. **Appendix** presents auxiliary data and proofs.

## 2. Mathematical model

### 2.1 CIR bridge

Our CIR bridge is a version of that originally proposed in Yoshioka [29] where drift and diffusion coefficients are allowed to be time-dependent. Time is denoted as $t \geq 0$. We consider a CIR bridge with the initial time 0 and a terminal time $T > 0$. Let $X = (X_t)_{0 \leq t \leq T}$ be unit-time fish count, e.g., number of fish counted in each time unit at a fixed point placed along a river. Approximating discrete fish counts with a continuous-time SDE would induce modeling errors, which may become severer particularly when Poisson-like discreteness may become non-negligible. Nevertheless, this approach is attractive because of its analytical tractability as demonstrated below.

The CIR bridge considered in this paper is the Itô's SDE:

$$\underbrace{\mathrm{d}X_t}_{\text{Increment}} = \underbrace{\left( \underbrace{a_t}_{\text{Acceleration}} - \underbrace{\frac{r}{T-t} X_t}_{\text{Reversion}} \right) \mathrm{d}t}_{\text{Drift}} + \underbrace{\sigma_t \sqrt{\frac{r}{T-t} X_t} \mathrm{d}B_t}_{\text{Diffusion}}, \quad 0 < t < T \tag{1}$$

subject to homogeneous initial and terminal conditions $X_0 = X_T = 0$, $B = (B_t)_{t \geq 0}$ is a 1-D standard Brownian motion (1.1 Definition in Karatzas and Shreve [57]), $r > 0$ is the reversion coefficient, $a = (a_t)_{0 \leq t \leq T}$ is a nonnegative acceleration coefficient, and $\sigma = (\sigma_t)_{0 \leq t \leq T}$ is a nonnegative volatility coefficient. When the unit of time is [T] (e.g., day or min), the unit of $X$ is [Count/T], and those of $a$, $r$, and $\sigma$ are [Count/T$^2$], [-], and [Count$^{1/2}$/T$^{1/2}$], respectively.

Each term of (1) is explained as follows. The acceleration term represents the unit-time supply of individual fish from downstream reaches of the observation point, and hence is considered external to the fish at the observation point. The reversion term represents the decay of fish count toward 0, which mathematically suppresses $X$ from explosion, and because of the factor $\frac{1}{T-t}$, enforces the terminal condition $X_T = 0$ with probability 1 under certain conditions (**Proposition 2**). This term plays a role in controlling the number of migrants so that it possibly becomes positive during the daytime only. The diffusion term represents fluctuation in fish count, and its nonlinearity, modulated by the coefficient $\sigma$, enables modeling both intermittent and non-intermittent sample paths within a single mathematical framework. This term conceptually describes the randomness of the number of migrants; however, it does not distinguish the factors causing the randomness (i.e., whether they are environmental or biological factors, etc.). From a biological standpoint, the singularity $\frac{1}{T-t}$ arises from the time change [29]

$$X_t \mathrm{d}\left( r\ln\left(\frac{1}{T-t}\right)\right) + \sigma_t \sqrt{X_t}\,\mathrm{d}\left( B_{r\ln\left(\frac{1}{T-t}\right)}\right) = \frac{r}{T-t} X_t \mathrm{d}t + \sigma_t \sqrt{\frac{r}{T-t} X_t}\,\mathrm{d}B_t,\quad 0 < t < T \tag{2}$$

in the sense of law. The function $r\ln\left(\frac{1}{T-t}\right)$ to be understood as auxiliary time is convex and blows up near the terminal time, representing a biological clock that drives fish migration so that the migration occurs only during daytime. Finally, note that a CIR bridge can be interpreted as an aggregation of agent-based models (e.g., [55]).

### 2.2 Moments and autocorrelation

Formal calculations that apply expectations $\mathbb{E}$ of $X_t$ and $X_t^2$ show that the average $\mathbb{E}\left[X_t\right]$ and variance $\mathbb{V}\left[X_t\right]$ satisfy the ordinary differential equations (ODEs)

$$\frac{\mathrm{d}}{\mathrm{d}t}\mathbb{E}\left[X_t\right] = a_t - \frac{r}{T-t}\mathbb{E}\left[X_t\right],\quad 0 < t < T \tag{3}$$

and

$$\frac{\mathrm{d}}{\mathrm{d}t}\mathbb{V}\left[X_t\right] = \frac{r\sigma_t^2}{T-t}\mathbb{E}\left[X_t\right] - \frac{2r}{T-t}\mathbb{V}\left[X_t\right],\quad 0 < t < T \tag{4}$$

subject to the initial conditions $\mathbb{E}\left[X_0\right] = \mathbb{V}\left[X_0\right] = 0$. Once the average $\mathbb{E}\left[X_t\right]$ and variance $\mathbb{V}\left[X_t\right]$ become available, we obtain the autocorrelation function

$$\mathrm{ACF}_{t,h} = \frac{\mathbb{E}\left[\left(X_t - \mathbb{E}\left[X_t\right]\right)\left(X_{t+h} - \mathbb{E}\left[X_{t+h}\right]\right)\right]}{\sqrt{\mathbb{V}\left[X_t\right]\mathbb{V}\left[X_{t+h}\right]}},\quad 0 < t \le t+h < T, \tag{5}$$

where $h$ represents the time lag. This autocorrelation function depends on the current time $t$ because the SDE (1) is not stationary. The ODEs (3) and (4) can be solved once the coefficients $a$, $r$, and $\sigma$ are specified. Our approach is in the opposite direction; we first specify $\mathbb{E}\left[X_t\right]$ and $\mathbb{V}\left[X_t\right]$ as time-dependent curves, and then find $a$ from (3) and $\sigma$ from (4) as coefficients depending on $r$. Finally, $r$ is identified by matching theoretical and empirical autocorrelation functions.

Applying the variation of constants formula to (1) yields

$$\begin{aligned} X_t &= \int_0^t a_s \left(\frac{T-t}{T-s}\right)^r \mathrm{d}s + \int_0^t \sigma_s \sqrt{\frac{r}{T-s}X_s}\left(\frac{T-t}{T-s}\right)^r \mathrm{d}B_s \\ &= \mathbb{E}\left[X_t\right] + \int_0^t \sigma_s \sqrt{\frac{r}{T-s}X_s}\left(\frac{T-t}{T-s}\right)^r \mathrm{d}B_s \end{aligned},\quad 0 \le t \le T, \tag{6}$$

where the second term is a martingale (see **Section A2**) and is denoted as $M_t$. The following **Proposition 1** shows a closed-form formula for the autocorrelation function $\mathrm{ACF}_{t,h}$.

***Proposition 1***

$$\mathrm{ACF}_{t,h} = \left(1-\frac{h}{T-t}\right)^{r} \sqrt{\frac{\mathbb{E}\left[M_t^2\right]}{\mathbb{E}\left[M_{t+h}^2\right]}}, \quad 0 < t \leq t+h < T . \qquad (7)$$

Now, we introduce our average and variance curves. For $0 \leq t \leq T$,

$$\mathbb{E}\left[X_t\right] = A\left(\frac{t}{T}\right)^{m}\left(1-\frac{t}{T}\right)^{n} \qquad (8)$$

and

$$\mathbb{V}\left[X_t\right] = \mathbb{E}\left[M_t^2\right] = V\left(\frac{t}{T}\right)^{p}\left(1-\frac{t}{T}\right)^{q}, \qquad (9)$$

where $A, V > 0$ are scale parameters that control the size of the moments and $m, n, p, q > 0$ are power parameters that modulate their functional shapes (see **Figure 1** with a normalization). With (9) in mind, we can rewrite (7) as

$$\mathrm{ACF}_{t,h} = \left(\frac{t}{t+h}\right)^{\frac{p}{2}}\left(1-\frac{h}{T-t}\right)^{r-\frac{q}{2}}, \quad 0 < t \leq t+h < T . \qquad (10)$$

Functional forms of (8) and (9) are motivated by the empirical finding that the average and variance of 10-min fish count data seem to be reasonably modelled as unimodal curves [29]; see also **Section 3**). Later, we show that these formulae apply to daily fish migration data as well. Finally, other functional forms of these coefficients would be possible in theory, while such a model may become less tractable than that used in this paper.

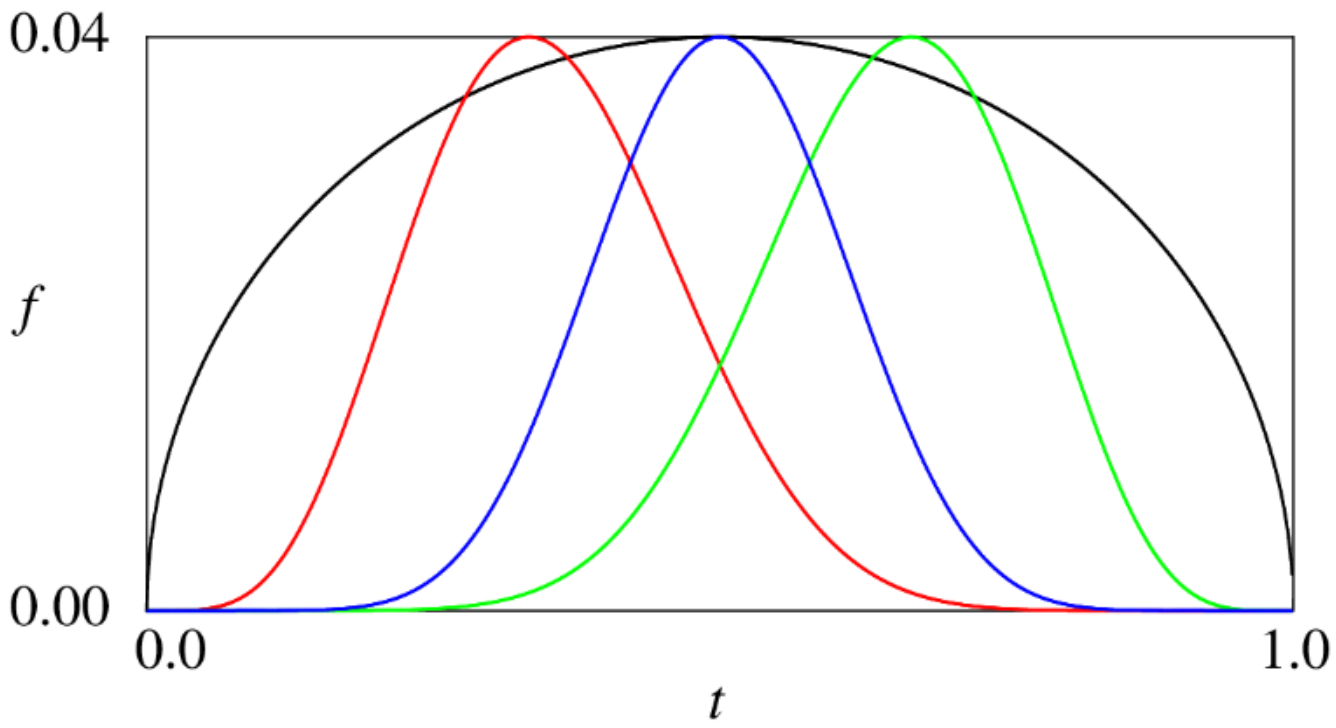

**Figure 1.** Functional shapes of $f_t = c_{m,n} t^m (1-t)^n$ ($m, n, c_{m,n} > 0$) with $c_{m,n}$ chosen so that $\max_{0 \le t \le 1} f_t = 1$. Values of $(m, n)$ are Black: $(0.5, 0.5)$, Red: $(5, 10)$, Green: $(10, 5)$, and Blue: $(10, 10)$. For the models using real data, see **Section 3**.

We conclude this subsection with an admissibility condition for values of $r,m,n,p,q$ because they turn out not to be arbitrary to obtain nonnegative $a$ and $\sigma^2$. First, substituting (8) into (3) yields

$$a_t = \frac{\mathrm{d}}{\mathrm{d}t}\mathbb{E}[X_t] + \frac{r}{T-t}\mathbb{E}[X_t] = \frac{A}{T^{m+n}}t^{m-1}(T-t)^{n-1}\left((r-m-n)t+Tm\right). \tag{11}$$

Similarly, substituting (8) and (9) into (4) yields

$$\sigma_t^2 = \frac{T-t}{r}\frac{1}{\mathbb{E}[X_t]}\left(\frac{\mathrm{d}}{\mathrm{d}t}\mathbb{V}[X_t] + \frac{2r}{T-t}\mathbb{V}[X_t]\right) = \frac{V}{Ar}T^{m+n-p-q}\left(t^{p-m-1}(T-t)^{q-n}\left((2r-p-q)t+Tp\right)\right). \tag{12}$$

Then, the last lines of (11) and (12) show the following admissibility condition, stating that the reversion $r$ needs to be sufficiently large to ensure the nonnegativity of $a$ and $\sigma^2$:

$$r \geq \max\left\{n, \frac{q}{2}\right\}. \tag{13}$$

We also have $0 \leq \mathrm{ACF}_{t,h} \leq 1$ under the condition (13). In this paper, we always assume the condition (13). We did not use any constrained optimization techniques for the data in this study except for that all the parameters are not negative.

The following **Proposition 2** is based on Yoshioka [29], stating that the SDE (1) under (13) admits a pathwise unique solution that is continuous and attains the initial and terminal conditions.

***Proposition 2***

*Assume (13). Then, the SDE (1) with (11) and (12) admits a pathwise unique solution that is continuous and nonnegative, and attains the initial and terminal conditions* $X_0 = X_T = 0$ *with probability 1.*

### 2.3 Feller condition

We introduce the Feller condition, which is vital in this study. To explain this condition, we introduce the Feller index $F_t$ whose value indicates whether the model generates intermittent sample paths or not:

$$F_t = \frac{1}{2}\frac{1}{a_t}\left(\sigma_t\sqrt{\frac{r}{T-t}}\right)^2 = \frac{\sigma_t^2}{2a_t}\frac{r}{T-t}, \quad 0 < t < T. \tag{14}$$

Intuitively, the Feller index $F_t$ as a nondimensional quantity evaluates the dominance of diffusion $\sigma_t\sqrt{\frac{r}{T-t}}$ relative to acceleration $a_t$, the source in (1), at each time $t$. The present Feller index is just a nonstationary version of that for the stationary one (e.g., Chapter 1 in Alfonsi [36]). A larger $F_t$ indicates higher dominance of diffusion, and vice versa, and sample paths of (1) become intermittent (resp., non-intermittent) at $t$ such that $F_t > 1$ (resp., $F_t \leq 1$). Therefore, the critical value of the Feller index is 1, and the distance from this critical value quantifies how much the model is (non-)intermittent. From a biological standpoint, one can judge whether the fish migration at an observation point is intermittent or not once the model is identified. As computationally discussed in the previous study [49], predicting intermittent CIR bridges becomes more difficult with coarser observations, even if the observation scheme is realistic.

Substituting (11) and (12) into (14) yields

$$F_t = \frac{V}{2A^2} T^{2m+2n-p-q} \frac{(2r-p-q)t+Tp}{(r-m-n)t+Tm} t^{p-2m} (T-t)^{q-2n}, \quad 0<t<T, \tag{15}$$

which increases with respect to the coefficient $VA^{-2}$, clearly showing the relationship between the (square of) coefficient of variation and Feller index according to (11) and (12). Moreover, the Feller index as a function of time has distinctive regularity conditions near $t=0,T$ depending on the values of $p-2m$ and $q-2n$. For the functional shapes in an application, see **Section 3**.

### 2.4 Modeling of intraday migration

The treatment of intraday fish migration data is the same as that proposed by Yoshioka [29], where times $0$ and $T$ are considered sunrise and sunset, respectively; both are deterministically found from meteorological data.

### 2.5 Modeling of daily migration

In contrast to the daily case, the initial and terminal times of migration need to be determined from data. Assume that migrants are counted at a fixed point and that we have their record for $N \in \mathbb{N}$ days (Day 1, Day 2, …, Day $N$). Assume further that in this record, the positive fish count is first and finally observed on Day $S$ (called the start day in the sequel) and Day $E$ (called the end day in the sequel), respectively, where $1 \le S \le E \le N$. The initial time is set as the beginning of Day $S$, and the terminal time is set as the end of Day $E$. The total migration duration is then estimated to be $D \equiv E-S+1$. The start and end days are different between years, even at the same observation point, and hence they are considered random variables that may depend on environmental conditions such as water temperature. For diadromous fish species, the increase in air temperature, which is highly correlated with water temperature, has been associated with earlier upstream migration [58]; autumn-spawning migratory fish species migrate earlier with warmer temperature [59]; the yearly increase in water temperature has been shown to delay the timing of spawning peaks and cessations of the spawning migration of Shad [60]; and the downstream spawning migration of Ayu have been reported to be triggered by low water temperature and increased streamflow discharge [61].

For modeling simplicity, we consider that what is random in the proposed model is the duration $D = E-S+1\ (=T)$ by a normalization $S=1$. In this formulation, only the terminal time is randomized. In our application, the length of $D$ is around 127 (see **Section A1**). We thus approximate the discrete-time dynamics by a continuous-time representation, enabling the use of the bridge to study the effective properties of the data. Moreover, applying the continuous-time framework to the daily count data allows for theoretical comparisons of fish migration phenomena on different timescales by a unified model.

Mathematically, the model in the daily case may be seen as a bridge with a random pinning time. Such bridges based on Brownian motion and Lévy processes have been studied extensively [62,63], whereas those of SDEs, particularly the proposed form, remain relatively underexplored. We do not delve deeply into the mathematical analysis of the proposed model with a random pinning time because it reduces to a bridge once the terminal time is specified. Indeed, with the proposed normalization method the start and end days each year can be transformed to nondimensional times 0 and 1, and the randomness of these days formally disappears (see **Section 3.2.2**).

## 3. Application

### 3.1 Target fish and study site

We apply the CIR bridge to fish count data of migrating juvenile Ayu at the Nagara River Estuary Barrage (5.4 km from the mouth of the Nagara River in Japan), which is a major river along which the fish migrate each year. Our focus in this study is the juvenile upstream migration of Ayu (**Figure 2**), which constitutes a key element in its one-year life cycle (e.g., Watanabe [64]) because the number of migrants that successfully arrive at midstream river reaches would critically influence their growth and reproduction success.

Nagara River Estuary Barrage is equipped with fishways, through which migratory fish species can pass, and an automatic video monitoring system has been installed at one of them since 2021[1]. Before that (2000 to 2020), migrating fish had been counted visually (i.e., manually) based on recorded video images. The data we use in this study are the fish count of juvenile Ayu monitored at the fishway each day (2003 to 2025, considering the data quality) and every 10 min (2023, 2024, and 2025) (**Figures 3 and 4**). The sunrise and sunset times have been found from the meteorological database[2]. The data used in this study do not include downstream migration data because no observations of downstream migration have been conducted at the study site.

**Section A1** presents the total fish count and the start and end of upstream fish migration for each year with some statistical analysis about migration duration. At the study site, the juvenile upstream migration of Ayu occurs from February to June each year. To the best of our knowledge, there are no or only a few cases in which both daily and intraday migration data are simultaneously available.

---

[1] https://www.water.go.jp/chubu/nagara/15_sojou/chousahouhou.html Last accessed on December 19, 2025.

[2] https://eco.mtk.nao.ac.jp/koyomi/dni/dni24.html Last accessed on December 19, 2025.

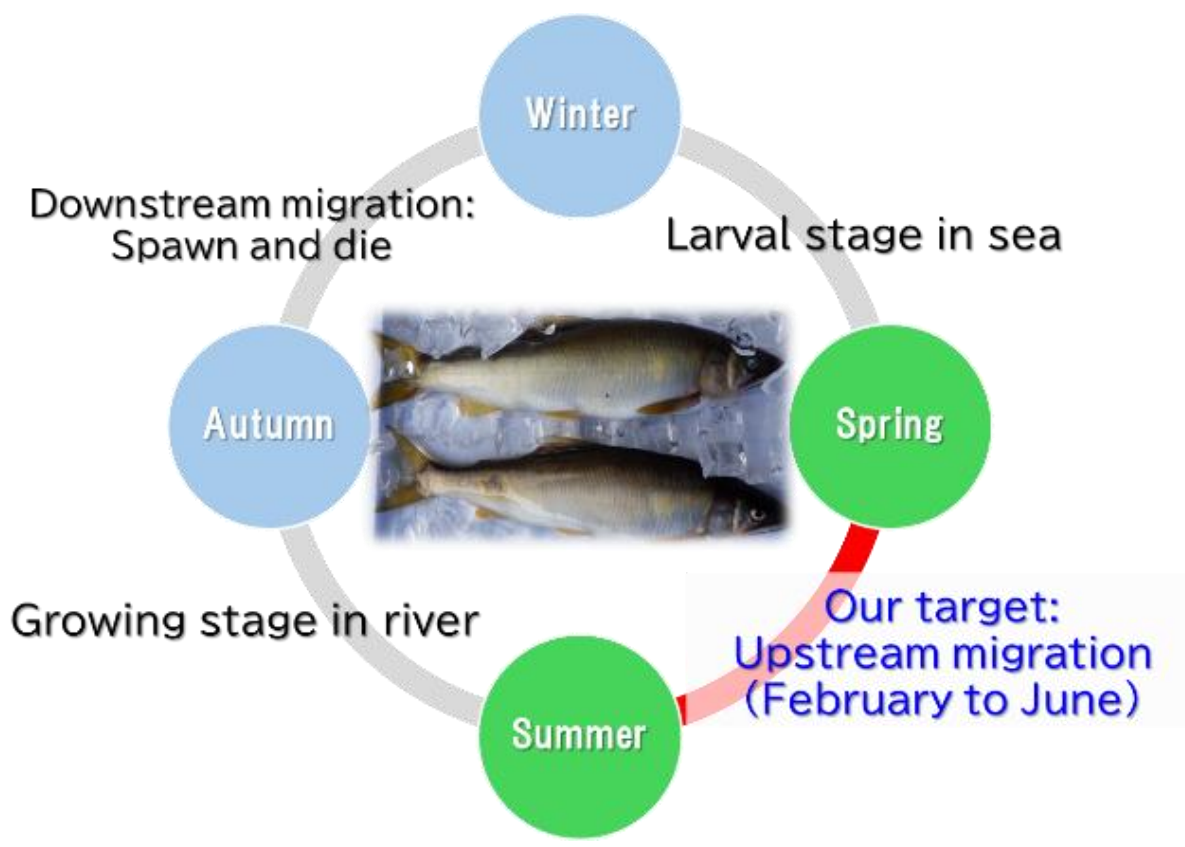

**Figure 2.** The target fish species Ayu.

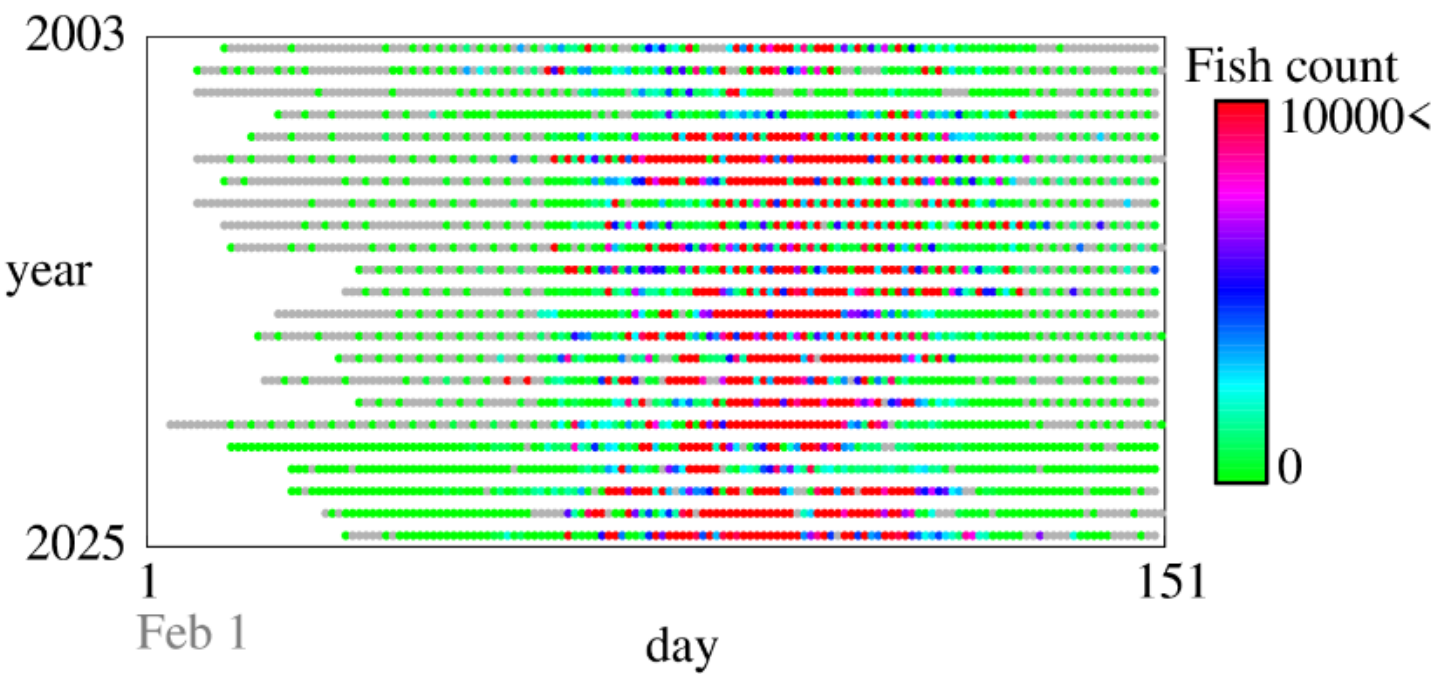

**Figure 3.** Daily fish count data of Ayu from 2003 to 2025. Grey circles represent zero count data, and blank areas represent no data.

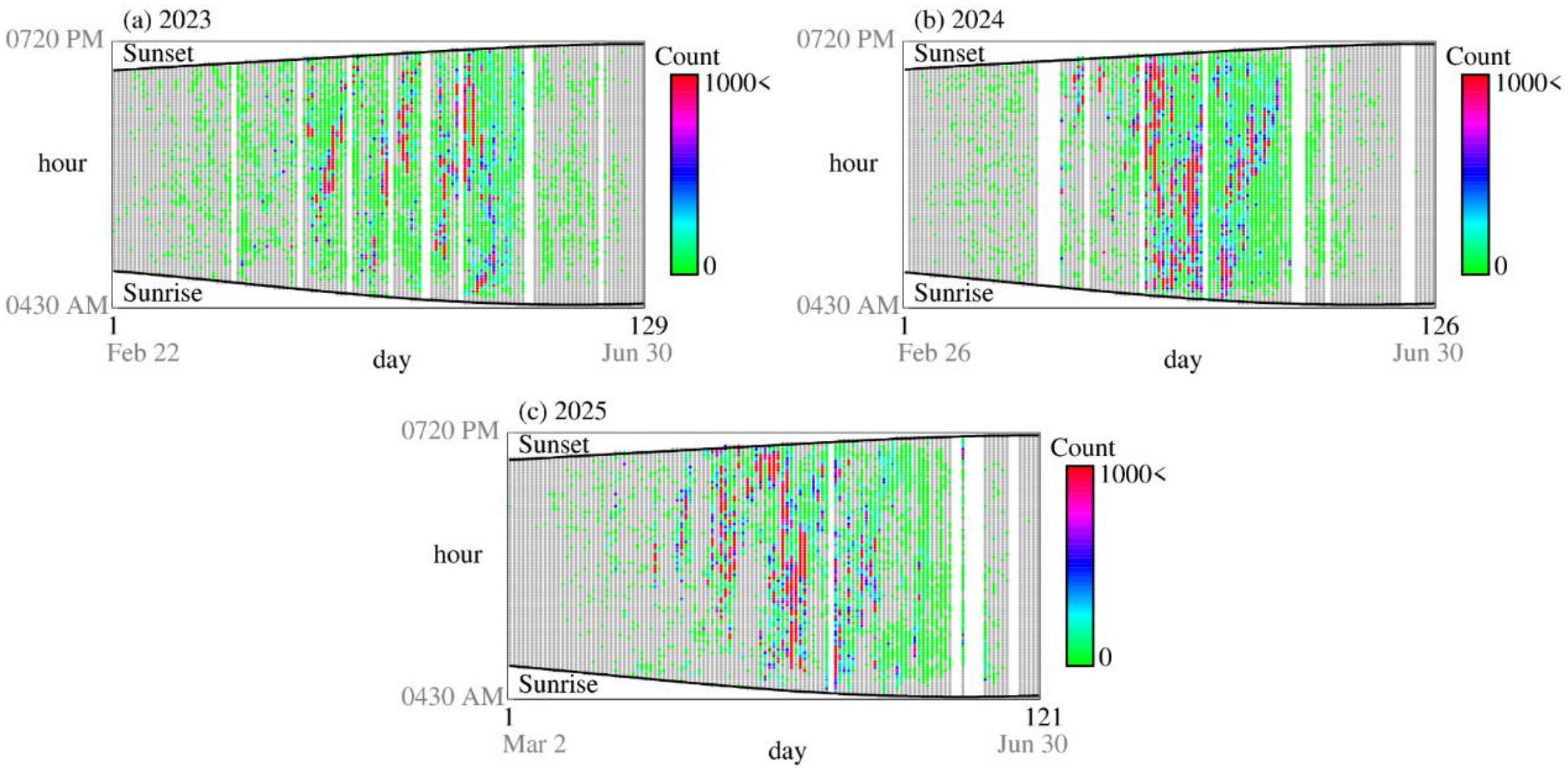

**Figure 4.** 10-min fish count data of Ayu. Grey circles represent zero count data, and blank areas represent no data. Sunrise and sunset times are also presented in the figure. The data plots have been presented in Yoshioka [55], and we have replotted them as new figure panels.

### 3.2 Data normalization and analysis

We identify the CIR bridge (1) for the daily and 10-min count data with normalizations explained below.

#### 3.2.1 Normalization of intraday fish count data

The procedure here is the same as that of Yoshioka [29], who normalized a CIR bridge with daytime duration, total daily fish count, and the unit time (10 min). We briefly explain this normalization procedure.

The sunset on some Day $N$ is expressed as $T_N > 0$ (recall that sunrise is at time 0), and the total number of migrants on this day as $Y_N$. The unit time (10 min) is $\Delta t > 0$. The SDE (1) on Day $N$ is given simply by the substitution $T = T_N$ by assuming that Brownian motions are independent of each other among different dates. Because the total fish count $Y_N$ is different for different days, we employ the following normalized variables: $s_N = \frac{t}{T_N} \in [0,1]$, $Z_{s,N} = \frac{\Delta t}{Y_N} X_{t,N}$, $\bar{a}_s = \frac{a_{t,N} T_N \Delta t}{Y_N}$, $\bar{r} = r_N$, $\bar{\sigma}_s = \sigma_{t,N} \sqrt{\frac{\Delta t}{Y_N}}$, and the subscript $N$ represents day. Here, following Yoshioka [29], we used an ansatz that $\bar{a}$, $\bar{r}$, and $\bar{\sigma}$ are independent of $N$. Substituting these quantities into (1) yields, in the sense of law,

$$\begin{aligned} \mathrm{d}\left(\frac{Y_N}{\Delta t} Z_{s,N}\right) = & \left(\bar{a}_s \frac{Y_N}{T_N \Delta t} - \frac{\bar{r}}{T_N - T_N s} \frac{Y_N}{\Delta t} Z_{s,N}\right) \mathrm{d}(T_N s) \\ & + \left(\bar{\sigma}_s \sqrt{\frac{Y_N}{\Delta t}}\right) \sqrt{\frac{\bar{r}}{T_N - T_N s} \frac{Y_N}{\Delta t} Z_{s,N}}\, \mathrm{d}\left(\sqrt{T_N} B_{s,N}\right) \end{aligned}, \quad 0 < s < 1, \tag{16}$$

which can be rewritten as

$$\mathrm{d}Z_{s,N} = \left(\bar{a}_s - \frac{\bar{r}}{1-s} Z_{s,N}\right) \mathrm{d}s + \bar{\sigma}_s \sqrt{\frac{\bar{r}}{1-s} Z_{s,N}}\, \mathrm{d}B_{s,N}, \quad 0 < s < 1 \tag{17}$$

with the initial and terminal conditions $Z_{0,N} = Z_{1,N} = 0$, where $B_{s,N}$ is a 1-D standard Brownian motion. The normalized SDE (17) can be seen as the original SDE (1) with $T = 1$. The average, variance, and autocorrelation of the unit-time fish count can be obtained under the same normalization $T = 1$. With these normalization procedures, we consider each daily 10-min unit time fish count time series as a sample path of the normalized SDE (17). Note that the Feller index $F_t$ does not depend on the normalization parameter $\Delta t$. In the sequel, we omit the bar " ¯ " from parameters and coefficients for simplicity.

#### 3.2.2 Normalization of daily fish count data

Because the total daily fish counts, namely, the total number of migrants, have significant yearly difference (**Section A1**), the data for each year are normalized as follows. The fish count on Day $N$ of Year $M$ is denoted as $Y_{M,N}$. The total number of migrants in Year $M$ is $\hat{Y}_M = \sum_{N=1}^{D_M} Y_{M,N}$, where $D_M$ is the duration of migration in Year $M$. The normalized daily fish count on Day $N$ of Year $M$ is set as $X_{M,N} = Y_{M,N} / \hat{Y}_M$, which satisfies $\sum_{N=1}^{D_M} X_{M,N} = 1$ by definition. Thus, the data for each year are discussed using the normalized fish count $X_{M,N}$.

Another normalization of the daily count data is employed because the total migration duration $D_M$ differs among the years. The migration duration is normalized to $1$, such that the time domain of migration of Year $M$ given by $(0, D_M)$ is linearly transformed to the unit interval $(0,1)$. We use the following formal normalization analogous to the intraday case: $s = \frac{t}{D_M} \in [0,1]$, $Z_s = \frac{\Delta t}{\hat{Y}_M} X_t$, $\bar{a}_s = \frac{a_{t,M} D_M \Delta t}{\hat{Y}_M}$, $\bar{r} = r_M$, and $\bar{\sigma}_s = \sigma_{t,M} \sqrt{\frac{\Delta t}{\hat{Y}_M}}$, where $\Delta t = 1$ (day) and the subscript $M$ represents year. With this normalization, each daily time-series data in a year can be considered as one sample path of the SDE (1). Again, the choice of $\Delta t$ does not affect the Feller index $F_t$.

The start $S$ and end days $E$ of migration summarized in **Table A1** have not been directly used in the model identification procedure in **Section 3.2.3**, and they do not affect the normalization procedure explained here. However, it is possible to conduct sensitivity analysis from a different perspective. For example, we can somehow study sensitivity of the model against data by artificially modulating $\Delta t = 1$ to distorted ones like $\Delta t \times (1 \pm \omega)$ with $\omega$ being a small value. In this case, the normalized drift coefficient will be scaled from $\bar{a}_s$ to $\bar{a}_s \times (1 \pm \omega)$ and the normalized volatility coefficient from $\bar{\sigma}_s$ to $\bar{\sigma}_s \times \sqrt{1 \pm \omega}$. The average and variance of fish count, and hance the parameters $A$ and $V$, are multiplied by $1 \pm \omega$ under these perturbations. In contrast, the parameters $m, n, p, q$ remain unchanged. Finally, according to (14), the Feller index $F_t$ does not change by these perturbations because $1 \pm \omega$ in the numerator and denominator cancel out. These observations imply that, under the assumed perturbations, there will be no great changes in the estimated parameter values. Of course, the above discussion addresses a limited range of data perturbations. Currently, we are working on building a robust CIR bridge based on data collection for fiscal year 2026 and field observations in other rivers (environmental DNA analysis, etc.).

### 3.2.3 Model identification

The CIR bridge is identified for each dataset (multiyear daily fish count from 2023 to 2025 and 10-min fish count from 2023 to 2025) based on the data processing method explained above. First, we fit parameters $A, m, n$ using the least-squares method between empirical and theoretical average curves using (3). Second, we fit parameters $V, p, q$ using the least-squares method between empirical and theoretical standard deviations ($\sqrt{\mathbb{V}[X_t]}$) using (4). Finally, we fit the remaining parameter $r$ using the least-squares method between empirical and theoretical autocorrelation functions using (10). The total number of empirical sample paths for computing these statistics are 23 for the daily case and 341 for the intraday case.

### 3.3 Results and discussion

**Table 1** summarizes fitted parameter values for intraday and daily cases. It also shows the corresponding normalized root mean-square errors (RMSEs), which refers to the RMSE divided by the theoretical average and variance integrated over the unit time interval $(0,1)$ for the average and variance. For autocorrelation function, the normalized RMSE refers to the RMSE divided by the theoretical one integrated over the triangular domain $0 \leq t+h \leq 1$ ($t,h \geq 0$). **Figures 5 and 6** compare empirical and fitted averages and standard deviations for each case, respectively. Similarly, **Figure 7** compares empirical and fitted coefficient of variations. **Figure 8** compares empirical and fitted autocorrelation functions.

First, the admissibility condition (13) is completely satisfied (**Table 1**). According to **Figures 5 and 6**, the theoretical average and standard deviation (containing equivalent information with variance) curves by the fitted models reasonably track the empirical ones. Comparison between the daily and intraday cases suggest their qualitative difference in the moments, such that both cases theoretically predict unimodal average and variance curves, while their shapes are qualitatively different between these cases. This difference is quantified in **Table 1**, where $m,n,p,q$ are around 10 for the daily case, while they are smaller than 1 for the intraday case. The average and variance (and hence standard deviation) are smoother in the daily case than the intraday case, and particularly those of the latter case are only Hölder continuous near the boundary, indicating a sharper, more abrupt onset/cessation of migration activity compared to smooth daily transitions on average. Their shapes are different between the two cases; profiles of the moments are mountain-like (neither convex nor concave) and concave for the daily and intraday cases, respectively, which is a significant difference between the two cases. This finding suggests that the acceleration $a$ of the fish count gradually (resp., suddenly) changes for the daily (resp., intraday) case.

**Figure 7** implies that the coefficient of variation is accurately predicted in the entire time interval for the intraday case, while accuracy degrades near the initial and terminal times for the daily case. However, this deviation between the empirical and theoretical coefficient of variations would not be critical in practice because there are only a few migrants during these time intervals. Interestingly, the coefficient of variation, which is dimensionless, is valued around 2 to 4 for both cases except near the initial and terminal times, which is a relevant common feature shared by the daily and intraday migration phenomena. This finding would be a hint toward more sophisticated modeling of upstream migration of Ayu in future, particularly that the upstream migration of this fish species is intermittent in both daily and intraday scales (see also the discussion about the Feller index presented later). The proposed CIR bridge successfully captures this remarkable property. **Figure 8** and **Table 1** imply that the autocorrelation decays fast in both cases, suggesting that bursts in the time series are rapidly suppressed. The fit is better in the intraday case than the daily case because of more fluctuating empirical autocorrelation function in the latter case. Fluctuations in the empirical autocorrelation function in the daily case are considered partly due to the smaller number of samples than in the intraday case.

**Figure 9** compares the Feller indices $F_t$ for each case, showing that for both daily and intraday cases, the fitted models predict the intermittency $F_t > 1$ ($0 < t < 1$), which is a similarity between the distinctive timescales. Therefore, it suggests the new finding that the juvenile upstream migration of Ayu at the study site fully violates the Feller condition both in daily and intraday timescales. The Feller index $F_t$ is extremely large near the initial and terminal times for the daily case, but there are almost no migrants around these time intervals. In both cases, $F_t$ is unbounded due to $p - 2m < 0$ and $q - 2n < 0$. The Feller index $F_t$ is smaller in the daily case than in the intraday case for moderate values of $t$ around the time 0.5, which corresponds to around April to May for the daily case and around noon for the intraday case. Intermittency is thus suggested to be stronger for the finer timescale in view of the Feller index as a nondimensional function of time. These findings about the Feller index, particularly its size exceeding the critical value 1, suggest that predicting fish migration phenomena is a challenging task.

For daily migration of other migratory fish species, literatures suggest that the intermittency of time series data of fish count seem to depend on species, years, and possibly observation points (e.g., Yeldham et al. [53] for shad and lamprey, Stewart et al. [65] and Xie et al. [66] for salmonids, and Weaver et al. [67] for eel). Therefore, a limitation of this study is the applicability of the mathematical framework to other fish species, and this point needs to be investigated in future. One difference between this and previous studies is that our data are based on the high-resolution and automatic count data, while many other studies are based on manual count or catch data, or acoustic count data. As discussed in Yoshioka [49], direct comparison between automatic and manual count data needs care because their accuracy may be significantly different from each other.

Finally, we quantify the difference in randomness inherent in the fish count in distinctive timescales. Sample paths of the CIR bridges are computed by the verified finite difference method of Yoshioka [29], which unconditionally preserves the nonnegativity of numerical solutions due to the structure-preserving discretization scheme [68]. The unit-time interval $(0,1)$ is divided into 50,000 equidistant subintervals, and we compute 1,000,000 sample paths for each computational case. **Figure 10** shows the computed sample paths of the unit-time fish count for each case. **Figure 11** shows the corresponding probability density functions (PDFs) at several time instances. The computed sample paths presented in **Figure 10** suggest that daily and intraday cases exhibit qualitatively distinctive behavior of the unit-time fish count. For the daily case, the fish count is almost zero near the initial and terminal times and the fish migration becomes active in intermediate time intervals, which is consistent with the empirical results. By contrast, for the intraday case, fish migration occurs intermittently except near the terminal time around which sample paths seem to be more oscillating and randomized, with bursts having much shorter durations. This finding corresponds to the asymmetric profiles (i.e., the center of gravity of plots is shifted to the right) of the theoretical average and standard deviation presented in **Figures 5(b) and 6(b)**.

According to **Figure 11(a)**, the daily case has a unimodal PDF that is peaked at zero and has a relatively heavy tail for intermediate times corresponding to the mountainous shape in the average and variance (**Figures 5(a) and 6(a)**). By contrast, the PDFs of the intraday case are almost the same between $t = 0.1$ and $t = 0.9$ (**Figure 11(b)**); indeed, the average and variance are not significantly varying except near the initial and terminal time, as shown in **Figures 5(a) and 6(a)**. A similarity between the two cases is the unimodal nature maximized at zero, which is qualitatively similar to that of the stationary CIR model at a stationary state violating the Feller condition (Chapter 1.2.3 in Alfonsi [36]). The variability of the PDF in time is therefore quantitatively different between the daily and intraday cases, while qualitatively the same due to both violating the Feller condition.

Consequently, our results suggest that the juvenile upstream migration of Ayu can be effectively described through the normalized CIR bridge violating the Feller condition on both daily and intraday timescales, and their difference appear as the size of Feller index and the parameter regime, i.e., regularity of the average and variance curves. This implies that the mathematical sciences developed for modeling the intraday fish migration, such as the minimization principle behind the migration dynamics [49] and mean-field descriptions [55], carry overs to the daily case with proper adaptations. A limitation in this study is that it does not directly give the microscopic collective behavior arising from self-organization (i.e., group size and shape during migration; Ioannou and Laskowski [32]). Its clarification would need finer fish count data, such as the arrival time of each migrant at the observation point, which is currently not available. We believe that the development of a proper agent-based model would become feasible with such data. Nevertheless, even if using such a microscopic model, analysis of the randomness and intermittency of intraday fish migration phenomena can be reduced to studying some statistical index like the Feller index. In this view, the present mathematical approach based on an SDE is a reasonable choice because it applies not only to the intraday data but also to the daily one.

**Table 1.** Estimated parameter values (Ave: average, Std: standard deviation, and ACF: autocorrelation function). All parameters are nondimensional.

| Parameter | | Daily | Intraday |
|---|---|---|---|
| $A$ | | 1.898E+04 | 2.056E-02 |
| $V$ | | 1.475E+04 | 3.034E-03 |
| $m$ | | 1.137E+01 | 4.146E-01 |
| $n$ | | 8.361E+00 | 1.954E-01 |
| $p$ | | 1.391E+01 | 3.635E-01 |
| $q$ | | 1.032E+01 | 1.395E-01 |
| $r$ | | 6.190E+01 | 5.683E+01 |
| Normalized RMSE | Ave | 3.874E-01 | 2.277E-01 |
| | Std | 5.302E-01 | 3.305E-01 |
| | ACF | 1.758E-01 | 5.472E-02 |

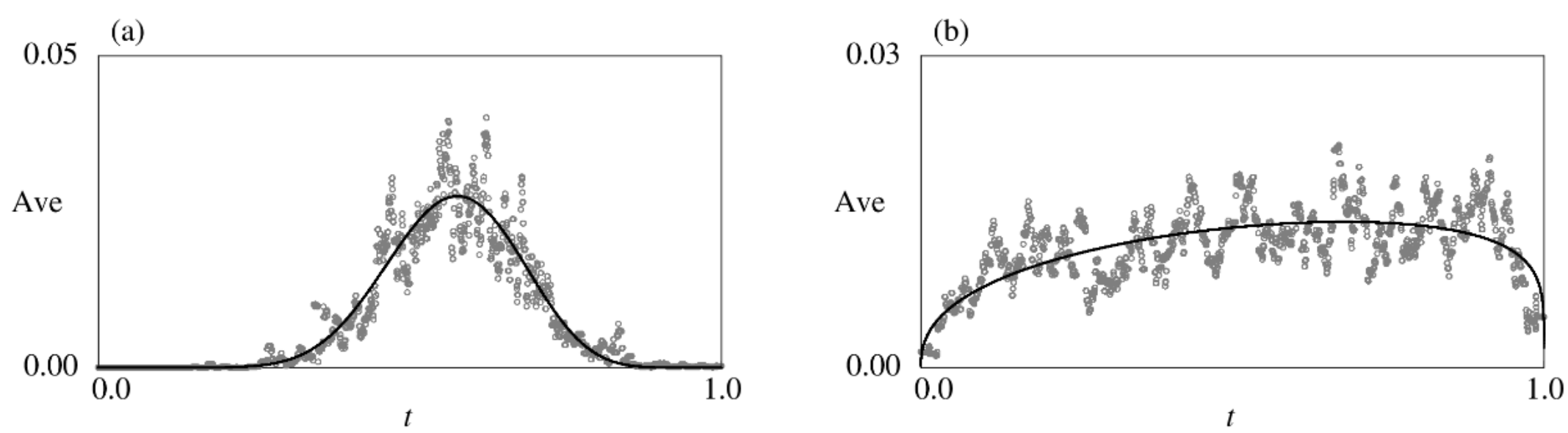

**Figure 5.** Comparison between empirical (circles) and fitted (curves) averages (Ave): (a) daily case (b) intraday case.

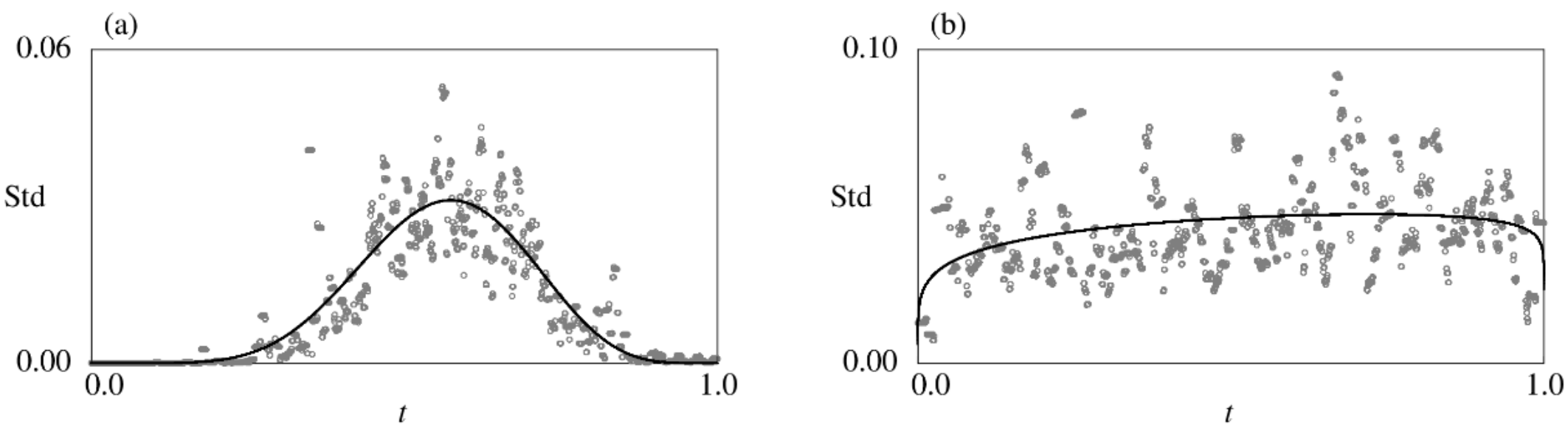

**Figure 6.** Comparison between empirical (circles) and fitted (curves) standard deviations (Std): (a) daily case (b) intraday case.

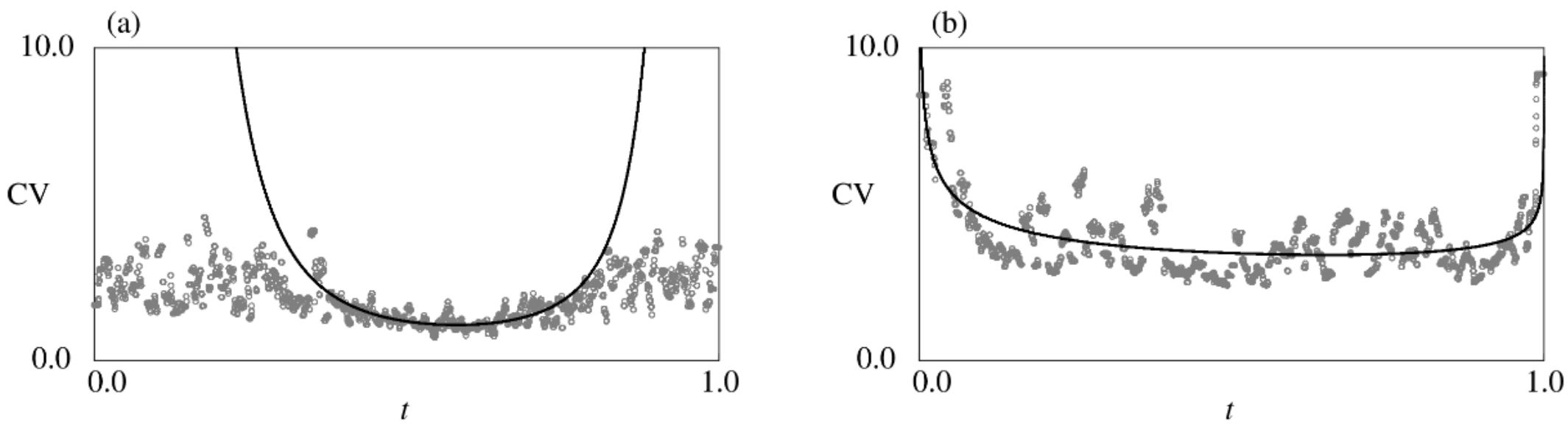

**Figure 7.** Comparison between empirical (circles) and fitted (curves) coefficient of variations (CV): (a) daily case (b) intraday case.

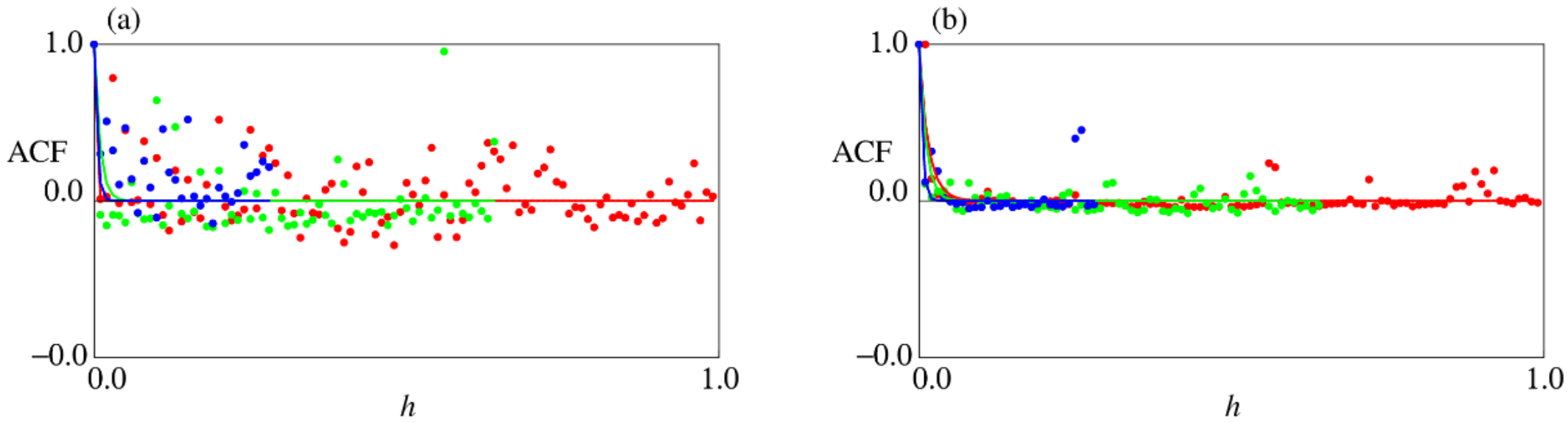

**Figure 8.** Comparison between empirical (circles) and fitted (curves) autocorrelation functions (ACF): (a) daily case (b) intraday case. Colors represent the results for $t = 0.01$ (red), $t = 0.36$ (green), and $t = 0.72$ (blue). These time instances have been chosen for representing early, mid, and late phases of the migration window.

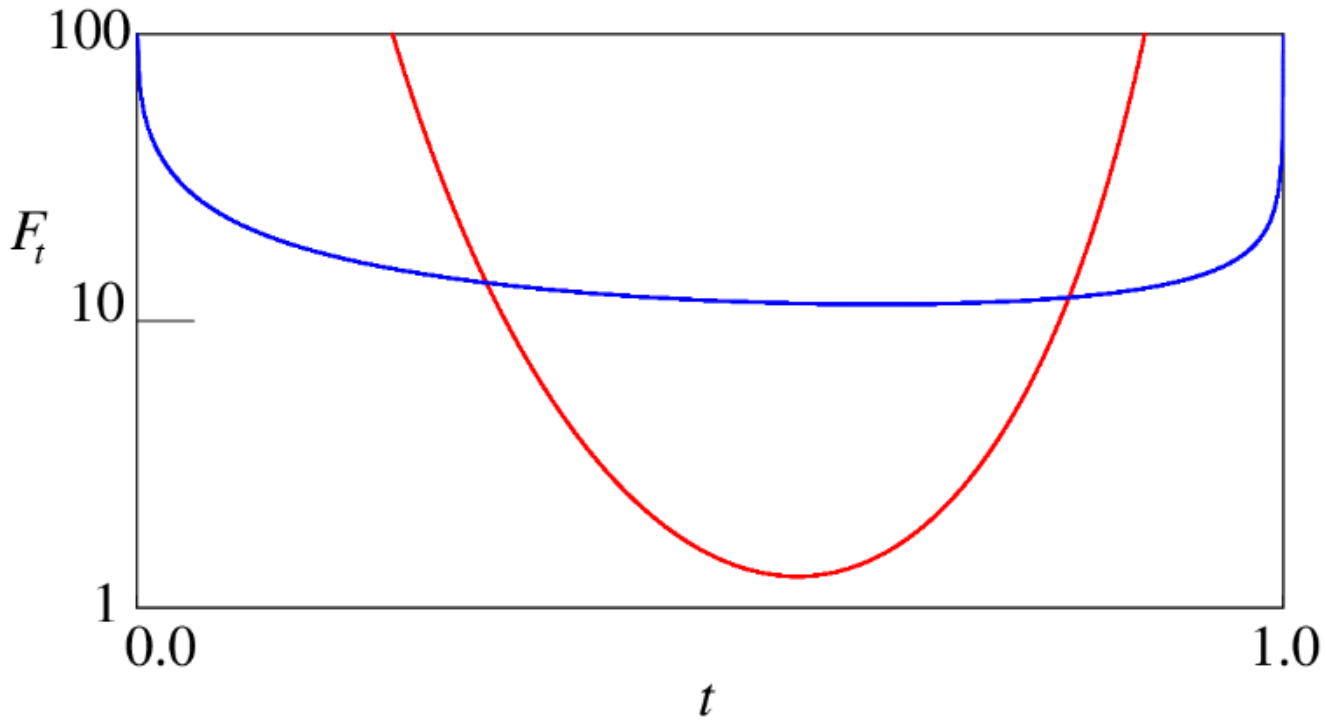

**Figure 9.** Comparison of the Feller indices $F_t$ between daily (red) and intraday (blue) cases on an ordinary logarithmic scale.

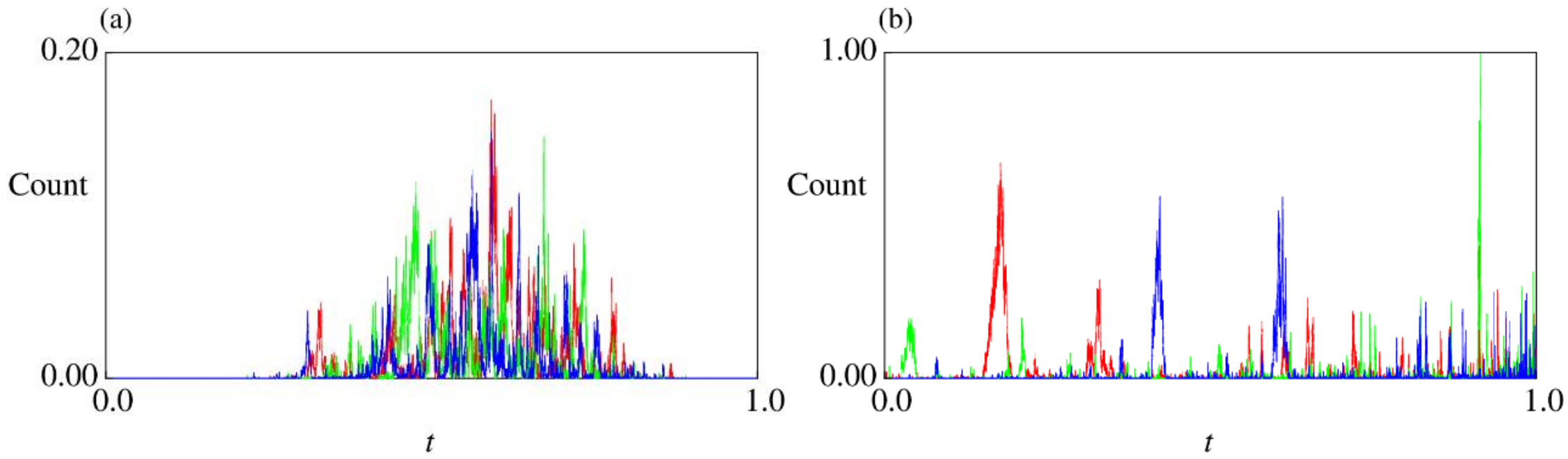

**Figure 10.** Computed sample paths of the unit time fish count: (a) daily case and (b) intraday case. Distinct colors (red, green, and blue) represent distinctive sample paths.

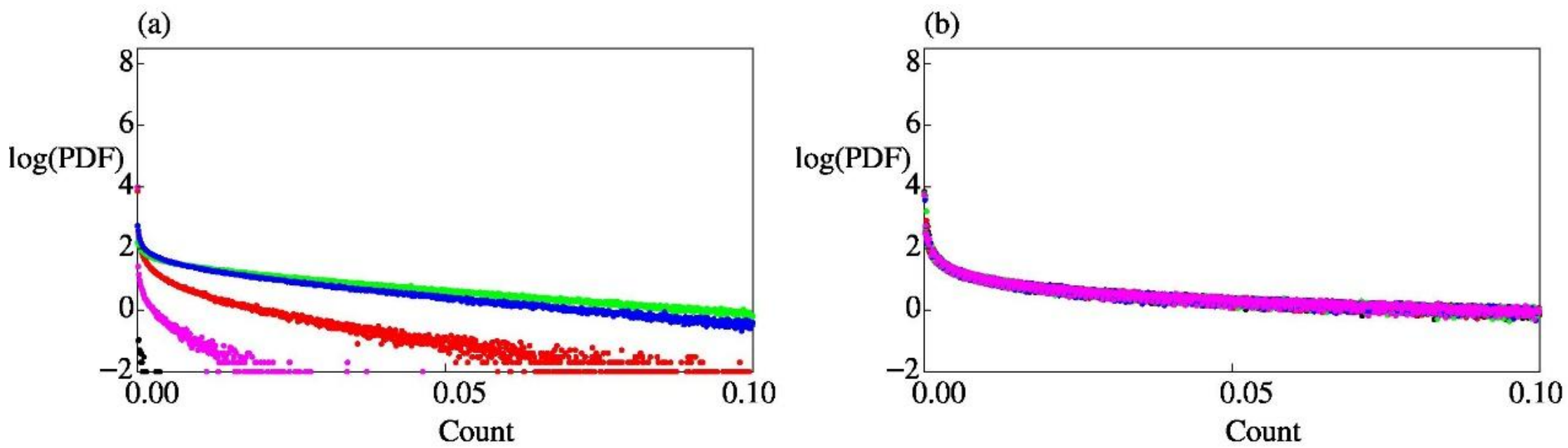

**Figure 11.** Computed PDFs of fish count at several instances: (a) daily case and (b) intraday case. Colors represent the results at times 0.1 (black), 0.3 (red), 0.5 (green), 0.7 (blue), and 0.9 (magenta).

## 4. Conclusion

We proposed a unified mathematical approach to analyze both daily and intraday fish migration phenomena. Daily and intraday fish counts at a study site were modeled using CIR bridges but with different parameter regimes. The analysis focusing on the Feller condition revealed that the sample paths show qualitatively distinctive properties depending on the relative size of randomness and drift. We found that the regularity of the coefficients of the CIR bridge is qualitatively different between the daily and intraday cases, which was reflected in the different parameter regimes in the average and variance curves. The results obtained in this study suggested that the Feller index, which is given by a nondimensional curve, can be used for evaluating the randomness and intermittency of fish migration across different timescales. Particularly, the Feller condition was violated both in daily and intraday timescales.

The results obtained in this study would deepen the understanding of the collective motion of a wide variety of migratory fish species. A natural question is the possibility of categorizing the biology and ecology of migratory fish species according to the Feller index, which possibly be able to assist modern empirical models for explaining fish migration such as otolith-based statistical model [69]. Fortunately, we could conduct this study because both daily and intraday fish count data were available at a study site; however, such data for fish species other than Ayu are still rare to the best of the author's knowledge. It is uncertain (at least for the author) whether similar migration data is available for other fish species, but the applicability of our approach to other fish species is an important research topic. We consider that how the proposed framework could be falsified or distinguished from alternative stochastic descriptions would depend on the data availability. Further data accumulation is considered necessary in this regard. Both technical progress and theoretical development are necessary to establish more detailed and realistic modeling of fish migration. Moreover, we did not account for the size spectrum in migrants, which is possibly a key factor in modeling migration because body size would affect the physiological performance of individual fish. We continue to study the biology and ecology of diadromous fish species, including Ayu, toward sustainable coexistence between nature and humans. The theoretical and computational results obtained in this study would effectively connect their coarse-scale (daily) and fine-scale (intraday) migration phenomena.

Finally, there is at least another timescale that was not dealt with in this study, which is the annual scale, because Ayu has a one-year life cycle. To approach the annual timescale in fish migration, we need to account for growth and reproduction driven by environmental changes such as river water temperature and discharge. The strength of the CIR bridge lies in its simplicity as a conceptual model. While it may be possible to develop models that take into account external environmental factors, there is a concern that these models would become much more complex and completely different from the model used in this study. This point will be addressed in future research.

## Appendix

### A.1 Auxiliary data

**Table A1** shows the total fish count and the start and end of the juvenile upstream migration of Ayu at the Nagara River Estuary Barrage. The annual duration of migration is presented in **Table A1**. The average, median, standard deviation, and skewness of the migration duration are 127.8 (day), 127 (day), 8.8 (day), and $-0.043$, respectively. Therefore, the upstream migration of Ayu at the study sites continues about four months each year. The standard deviation is much smaller than the average because the coefficient of variation of the migration duration is 0.068. Moreover, the size of the skewness of migration duration is significantly smaller than 1. In these views, a possible simplest modeling strategy to generate yearly migration phenomena in the daily time scale maybe to model the migration duration as a uniform distribution with some prescribed minimum and maximum values (111 and 143 according to **Table A1**), which can be seen as a diffusion bridge with a random pinning time [62,63]. Our normalization method treats the fish counts for each year equally, so it can be interpreted that the relative numbers of fish counts for each year are adequately incorporated into a dimensionless variable.

**Table A1.** Total fish count and the start and end of juvenile upstream migration of Ayu at the Nagara River Estuary Barrage. The duration of migration each year is also presented.

| Year | Total fish count | Start day | End day | Duration |
|---|---|---|---|---|
| 2025 | 1,318,281 | 2-Mar | 28-Jun | 119 |
| 2024 | 1,236,102 | 28-Feb | 26-Jun | 120 |
| 2023 | 852,596 | 22-Feb | 28-Jun | 127 |
| 2022 | 224,397 | 22-Feb | 30-Jun | 129 |
| 2021 | 403,459 | 13-Feb | 30-Jun | 138 |
| 2020 | 812,342 | 13-Feb | 30-Jun | 139 |
| 2019 | 592,439 | 4-Mar | 28-Jun | 117 |
| 2018 | 847,565 | 21-Feb | 28-Jun | 128 |
| 2017 | 1,171,928 | 1-Mar | 24-Jun | 116 |
| 2016 | 702,028 | 17-Feb | 30-Jun | 135 |
| 2015 | 957,706 | 10-Mar | 28-Jun | 111 |
| 2014 | 608,661 | 5-Mar | 30-Jun | 118 |
| 2013 | 993,089 | 4-Mar | 30-Jun | 119 |
| 2012 | 590,157 | 13-Feb | 29-Jun | 138 |
| 2011 | 841,043 | 24-Feb | 30-Jun | 127 |
| 2010 | 471,415 | 25-Feb | 30-Jun | 126 |
| 2009 | 2,174,478 | 12-Feb | 30-Jun | 139 |
| 2008 | 2,695,955 | 13-Feb | 28-Jun | 137 |
| 2007 | 785,887 | 16-Feb | 30-Jun | 135 |
| 2006 | 130,024 | 20-Feb | 29-Jun | 130 |
| 2005 | 70,157 | 26-Feb | 29-Jun | 124 |
| 2004 | 315,018 | 8-Feb | 29-Jun | 143 |
| 2003 | 437,693 | 12-Feb | 16-Jun | 125 |

**A2. Proofs**

***Proof of Proposition 1***

A direct calculation shows

$$
\begin{aligned}
&\mathbb{E}\left[\left(X_t-\mathbb{E}[X_t]\right)\left(X_{t+h}-\mathbb{E}[X_{t+h}]\right)\right] \\
&=\mathbb{E}\left[\left(\int_0^t \sigma_s\sqrt{\frac{r}{T-s}X_s}\left(\frac{T-t}{T-s}\right)^r \mathrm{d}B_s\right)\left(\int_0^{t+h} \sigma_s\sqrt{\frac{r}{T-s}X_s}\left(\frac{T-t-h}{T-s}\right)^r \mathrm{d}B_s\right)\right] \\
&=\mathbb{E}\left[\left(\int_0^t \sigma_s\sqrt{\frac{r}{T-s}X_s}\left(\frac{T-t}{T-s}\right)^r \mathrm{d}B_s\right)\left(\int_0^{t} \sigma_s\sqrt{\frac{r}{T-s}X_s}\left(\frac{T-t-h}{T-s}\right)^r \mathrm{d}B_s\right)\right] \\
&=(T-t)^r(T-t-h)^r\,\mathbb{E}\left[\int_0^t \sigma_s^2\frac{r}{T-s}X_s\left(\frac{1}{T-s}\right)^{2r}\mathrm{d}s\right] \\
&=\left(1-\frac{h}{T-t}\right)^r\mathbb{E}\left[\int_0^t \sigma_s^2\frac{r}{T-s}X_s\left(\frac{T-t}{T-s}\right)^{2r}\mathrm{d}s\right] \\
&=\left(1-\frac{h}{T-t}\right)^r\mathbb{E}\left[M_t^2\right]
\end{aligned}
\quad , \tag{18}
$$

where we used the isometry (e.g., Proposition 4.12 in Privault [70]), and

$$
\sqrt{\mathbb{V}[X_t]\mathbb{V}[X_{t+h}]}=\sqrt{\mathbb{E}\left[M_t^2\right]\mathbb{E}\left[M_{t+h}^2\right]}. \tag{19}
$$

Combining (18) and (19) along with (5) yields (7).

□

***Proof of Proposition 2***

In view of Proof of Proposition 1 in Yoshioka [29], it suffices to show that

$$
M_t=\int_0^t \sigma_s\sqrt{\frac{r}{T-s}X_s}\left(\frac{T-t}{T-s}\right)^r \mathrm{d}B_s \tag{20}
$$

is a martingale for $0\le t<T$ and satisfies $\mathbb{E}\left[M_{T-\varepsilon}^2\right]\to 0$ as $\varepsilon\to 0$, but they follow from $\mathbb{V}[X_{T-\varepsilon}]=\mathbb{E}\left[M_{T-\varepsilon}^2\right]$ for any $\varepsilon\in(0,T)$ along with the formula (9) and the martingale convergence theorem (Theorem 27.3 of Jacod and Protter [71]).

□